# Thermodynamics of the ultrafast phase transition of vanadium dioxide

Shreya Bagchi[1], Ernest Pastor[2,3], José Santiso[4], Allan S. Johnson[1+], Simon E. Wall[4]

1. IMDEA Nanoscience, Calle Faraday 9, 28049 Madrid, Spain

2. CNRS, Univ Rennes, Institut de Physique de Rennes, Rennes, France.

3. CNRS, Univ Rennes, DYNACOM (Dynamical Control of Materials Laboratory) - IRL2015, The University of Tokyo, Tokyo, Japan.

4. Catalan Institute of Nanoscience and Nanotechnology (ICN2), CSIC and BIST, Campus UAB, Bellaterra, 08193 Barcelona, Spain

5. Department of Physics and Astronomy, Aarhus University, Aarhus, Denmark

+Corresponding author: allan.johnson@imdea.org

**Ultrafast photoexcitation is an emerging route to selective control of phase transitions. However, it is difficult to determine which modes govern the transformation and how effectively they are targeted by photoexcitation. This is exemplified in vanadium dioxide, which transitions from a monoclinic insulator to a rutile metal upon heating or photoexcitation. There is a long-standing debate about whether this transition is electronically or structurally driven and whether the structural component is coherent, driven by a single structural mode or thermal in nature. In this work, we develop a simple thermodynamic framework based on temperature-dependent ultrafast pump–probe measurements and contrast it to microscopic-detail-free modelling to identify the driving mechanism of the transition, revealing that population of the full thermal phonon spectrum, especially high-frequency oxygen modes, is necessary to stabilize the metallic phase. Our approach can straightforwardly be applied to determine the nature of other photoinduced phase transitions without the need for complex multi-messenger experiments and can guide new control strategies, even for incoherent transitions.**

Control over complex materials is essential for quantum technologies leveraging quantum properties[1–4]. One powerful emerging strategy is ultrafast photoexcitation, where light pulses can be tuned to drive specific modes or degrees of freedom, enabling selective control of phase transitions[3,5]. The driving field exerts a uniform and coherent force over large regions of the medium, potentially leading to homogenous and coherent transitions, while the use of ultrafast pulses permits high field strengths to be applied to materials without introducing damage and reduces energy re-distribution out of the desired mode.

The problem is that, in general, it is difficult to determine *a priori* which are the key modes that govern a given phase transition and that must be selectively excited, as well as whether the ultrafast photoexcitation exclusively impacts those modes. A growing body of work suggests that multimode couplings can induce anomalous signatures of coherent optical control[6,7], leading to deviations from the behaviour expected when a single mode is selectively driven. These findings highlight that the contributions from disorder and entropy are critical even at the ultrafast timescale[8–11].

As such, the mechanisms by which photoexcitation drives phase transitions on the ultrafast timescale remain contentious.

**Electronic and structural complexity**

These complications are exemplified in vanadium dioxide ($VO_2$)[12,13]. At room temperature, $VO_2$ is a monoclinic $M_1$-phase insulator with four vanadium ions per unit cell. On heating above the critical temperature of 343 K or upon photoexcitation with energy above the critical fluence, $VO_2$ undergoes a first-order structural phase transition to a metallic rutile R-phase with only two vanadium ions per unit cell (**Fig 1(a)**). The nature of the phase transition in $VO_2$ has been debated since its discovery[14], stemming from the complex interplay between electronic and structural changes that occur near the transition temperature. While some studies emphasize the role of electron correlation effects driving a Mott-like insulator-to-metal transition[15], others highlight the importance of lattice distortions characteristic of a Peierls-type structural transition[16], while more recently others have emphasized the role of entropic structural fluctuations[17,18].

When $VO_2$ is excited below the critical fluence, coherent phonons emerge within the monoclinic ($M_1$) phase[13,19–24], modulating the vanadium dimerization and tilt. Early studies proposed that this coherent motion persists across the transition[13], and attributed the phase change to these phonons of well-defined frequency and wave vector[21,25]. Conversely, the observation of an electronic transition with simultaneous coherent $M_1$ phonons was interpreted as evidence that photoexcitation melts the electronic Mott ordering independently of the resulting structural response[20,23]. Reports of emergent non-equilibrium phases in $VO_2$ under similar excitation conditions further support a decoupled mechanism[26–29].

In contrast, recent time-resolved diffuse X-ray scattering experiments[8,30] have challenged the relevance of the coherent-phonons, revealing rapid disordering of the low-temperature structure during the transition. The diffuse scattering intensity was found to increase within ~50 fs of photoexcitation, accompanied by an equally fast reduction in the Bragg peak intensity of the monoclinic phase, providing evidence of an order–disorder transition analogous to the temperature-driven one[17]. This interpretation was further corroborated by double-pump measurements[29], which showed that coherent excitation of the 6 THz Raman $A_{1g}$ mode had no effect on the energy required to initiate the structural phase transition. In contrast, the total energy required was modulated by the excitation of incoherent phonons. Notably, however, coherent motion at particular Bragg reflections persisted even in the presence of the disorder[8], suggesting a more complex interplay.

Such investigations in the time domain are complicated by the strong electron-phonon coupling leading to nearly simultaneous structural and electronic transitions on the sub-100 fs timescale[31], below the time-resolution of many probe methods. Moreover, the static heating, finite penetration depths[28], and the complexity of modelling such experiments in a highly-correlated material like $VO_2$, further complicate elucidating the role of phonons in the phase transition.

Here, we propose an alternative method to identify the key modes that govern the photoinduced phase transition in $VO_2$. We leverage the dimensional and particle statistical dependence of the ultrafast heat capacity to discern whether the photoinduced phase transition requires exciting only the fermionic electron bath in a purely electronic process[24] (**Fig 1(b)**), a few discrete bosonic phonon modes along specific lattice coordinates, corresponding to a low-dimensional coherent structural response[19,31] (**Fig 1(c)**), or the full *N*-dimensional bosonic phonon bath, in which many vibrational degrees of freedom across the lattice are populated in an entropic or thermal process[8] (**Fig 1(d)**). A thermodynamic treatment of each of these cases leads to a different temperature dependence of the critical fluence[22,26,28,29], which is independent of the microscopic details of the material under study, and can thus be applied to a complex system like VO$_2$.

Our results, based on the systematic analysis of the critical energy or fluence required to initiate the structural phase transition in VO$_2$ as a function of temperature, show that the transition is indeed of order-disorder or thermal type. Furthermore, we find an unexpectedly strong dependence on the population of the high-frequency oxygen phonon modes, which are often neglected in the discussion of VO$_2$. In contrast, the coherent Raman-active modes remain temperature independent, confirming their spectator role in the phase transition. Finally, we discuss how our method can be used to identify the controllability and key modes of light-induced phase transitions in quantum materials generally, and to define new control strategies even for incoherent transitions.

**Ultrafast Heat Capacity of VO$_2$**

Full-dimensional modelling of VO$_2$ has proven to be a particularly challenging task, and as a result phenomenological or toy models are often applied to understand experimental features in ultrafast experiments[31–34]. While these models help understand microscopic effects, the numerous approximations required make definitive assignments challenging. Here, instead, we apply an agnostic statistical thermodynamic treatment, which has the advantage of predicting characteristic scaling laws without the need for a microscopic description of the material.

The underlying idea is this: if one particular mode or degree of freedom is responsible for the ultrafast phase transition, the energy required to drive the phase transition depends only on that degree of freedom. Because, at a given temperature, there will always be a number of pre-existing excitations (quasiparticles) in the relevant modes given by the heat capacity of those modes, the total energy required to drive the transition will vary with the temperature depending on the heat capacity of those relevant modes, This is regardless of whether the photoexcitation itself is thermal or not. Previous measurements of VO$_2$ have shown that the energy required to initiate the structural transition is indeed temperature dependent, though these findings have been interpreted in contrasting ways across different studies[20,23,28] and this dependence has not been systematically analyzed.

In general the energy required to drive the transition can be given by $L - U(T)$, where $L$ is the total energy of the transition, including any latent heat or zero-point energy, and $U(T)$ is the relevant internal energy of the relevant mode prior to photoexcitation. For VO$_2$ we consider three different potential driving mechanisms with correspondingly different internal energy dependencies:

**Electronic transition:** For an electronic-only Mott-like transition, the photoexcitation creates electron-hole pairs, which rapidly equilibrate[35]. The electrons and holes can be described as a three-dimensional free-electron gas following Fermi statistics, and the electronic heat capacity arises from the thermal excitation of conduction electrons near the Fermi surface and is expressed as: $C_{el} = \gamma T$, where $\gamma$ is the electronic heat capacity coefficient determined by the density of available electronic states at the Fermi level and the Boltzmann constant (**Fig 1(b)**).

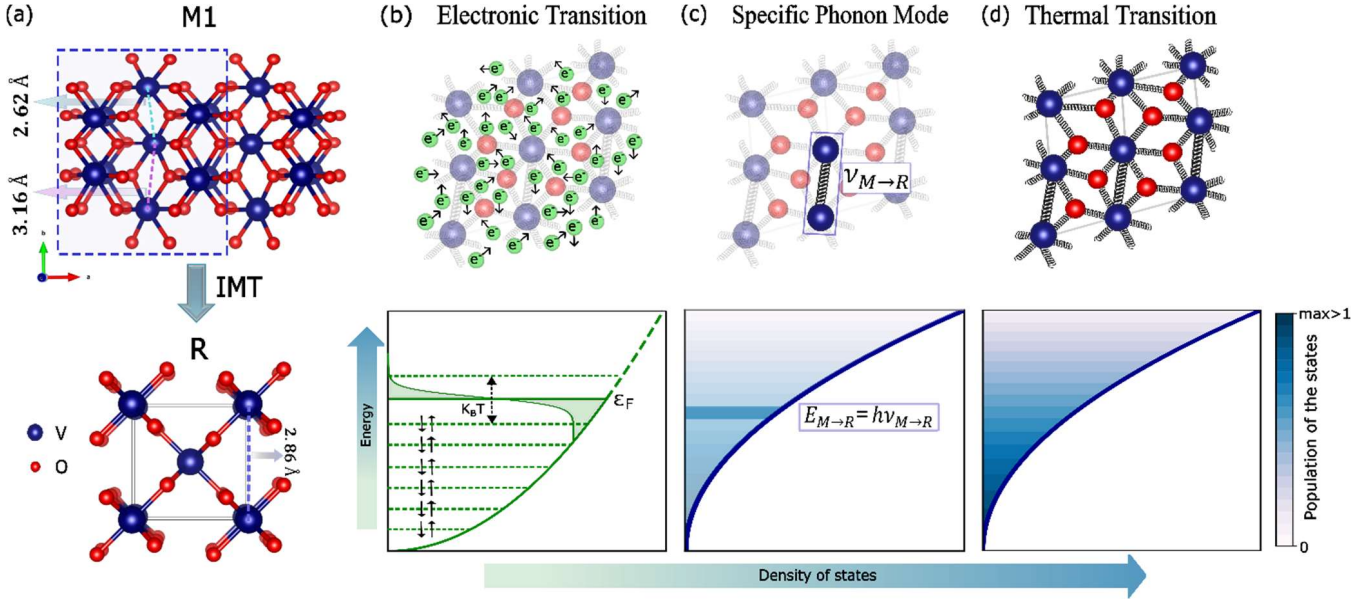

**Fig 1 Ultrafast Heat Capacity of VO$_2$:** (a) Crystal structure of VO$_2$ in the insulating monoclinic (M1) phase and the rutile metallic (R) phase. In the monoclinic phase, the V–V chains are dimerized, exhibiting alternating V–V distances of 3.16 Å and 2.62 Å, whereas in the rutile phase the V–V spacing is uniform at 2.86 Å. (b–d) Schematic real space (top panels) and density-of-states diagrams (bottom panels) depicting three distinct scenarios describing the insulator-to-metal transition (IMT) in VO$_2$: (b) a purely electronic mechanism, in which the population of excited carriers alone drives the collapse of the band gap. The electronic states obey fermionic occupation rules, and carriers can be thermally promoted to within a few $k_BT$ of the Fermi energy prior to photoexcitation. (c) specific phonon mediated pathway, where excitation of a particular lattice vibrational mode $\nu_{M \to R}$ directly breaks the V–V dimerization. Only energy deposited into this single bosonic mode controls the phase transitions, with all other modes acting as spectators. (d) full lattice thermalization, in which the necessary energy is distributed among all the bosonic phonon modes. The transition occurs when the cumulative vibrational energy (thermal and photoinjected) raises the lattice temperature above the critical temperature.

The internal energy contribution of electrons is $U_{el} = \int_0^T C_{el}(T')dT' \Rightarrow U_{el} = \gamma \frac{T^2}{2}$ and an electronic-only transition should thus show a quadratic dependence on the initial temperature. We note that in practice, the bandgap in the insulating phase means the temperature dependence is extremely weak, as very few carriers are thermally excited across the bandgap below the transition temperature, but this will not factor into our analysis and is captured within the $\gamma$ factor.

**Specific phonon modes:** For the case in which the transition is triggered by excitation of only one phonon mode $\nu_{M \to R}$ which links the high and low temperature phases, we can consider only the thermal population of that singular mode as given by the canonical partition function (**Fig 1(c)**). This is formally equivalent to the Einstein model, which treats the heat capacity as originating from phonons with a single characteristic frequency. While conventionally the Einstein model considers that all atoms vibrate at this frequency, because phonons obey Boson statistics, the existence of other modes - which do not directly contribute to the ultrafast heat capacity - does not modify the temperature dependence of the population of those modes that do. The internal energy of the phonons is given by $U_{ph}^E = U_0 \frac{\theta_E}{e^{\theta_E/T}-1}$, where $\theta_E = \frac{h\nu}{k_B}$ is the Einstein temperature, $\nu$ is the mode frequency, and $U_0$ represents the overall energy scale determined by the number of oscillators in the system. This energy thus depends solely on the frequency of the relevant mode and the temperature.

**Thermal transition:** In the thermal transition, the absorbed energy rapidly redistributes and equilibrates among all degrees of freedom. For the thermal transition in $VO_2$[36,37] and in many materials, the heat capacity is dominated by the lattice (phonon) contribution. The internal energy is then given by the integral over the phonon density of states weighted by the canonical partition function for Bosons, with the challenge then to select an appropriate density of states. The simplest model for the lattice contribution to the heat capacity is the Debye model, which treats the solid as an elastic continuum supporting vibrations up to a maximum cutoff frequency, the Debye frequency $\omega_D$ or correspondingly the Debye temperature $T_D$. (**Fig 1(d)**). The internal energy in this model is then $U_{ph}^D = 9NK_BT(\frac{T}{T_D})^3 \int_0^{T/T_D} \frac{x^3}{e^x-1} dx$. While the Debye model does not fully model the density of states used in any real material, it successfully captures the thermal behaviour of many solids and describes the populations of a continuum of bosonic states. Notably, however, the ultrafast phase transition in $VO_2$ occurs under constant volume[28] and involves primarily high-frequency optical phonons, with low-frequency acoustic modes contributing negligibly, if at all, to the transformation. Consequently, we instead adopt a gapped Debye model[38], which excludes the low-frequency acoustic phonons and accounts only for the higher-frequency optical modes. This is implemented by introducing a lower cutoff frequency (or temperature $T_c$) in the Debye integral. The internal energy in the gapped Debye model is then given by $U_{ph}^{GD} = \frac{9NK_BT^4}{{T_D'}^3 - T_c^3 + T_D' T_c^2 - T_c {T_D'}^2} \int_{T_C/T}^{T_D'/T} \frac{x(x-x_c)^2}{e^x-1} dx$.

Each of the three cases thus exhibit different characteristic temperature dependences with minimal consideration of the microscopic description.

**Experimental Results**

To investigate the energy required to trigger the structural phase transition we measured the fluence dependence of the phase transition at a wide range of temperatures below the critical temperature. We use ~12 fs duration 700 nm pump pulses to initiate the photoinduced phase transition in 60 nm epitaxial *c*-axis oriented $VO_2$ films grown by Pulsed laser deposition on $TiO_2$ (001) single crystal substrates (following the conditions described in ref. 39) and held in a close-cycle helium cryostat. The resulting changes in the reflection were probed with ~11 fs 1800 nm pulses; further details of the experimental setup are provided in the Methods and in Ref [40].

**Fig 2(a)** shows a representative high-resolution temporal trace obtained at 77 K for several pump fluences, which is broadly consistent with previous studies on $VO_2$[19–23,25]. At low fluences, we observed oscillations at 6 and 10 THz, corresponding to coherent phonon oscillations resulting from displace excitation of the $A_{1g}$ phonon modes. These oscillations exhibit a strong dependence on excitation fluence, with higher fluences leading to more rapid damping. Beyond 3 ps, no discernible oscillatory components are detected across the entire temperature range investigated, indicating complete dephasing of the coherent phonon modes irrespective of fluence.

In addition to the oscillatory component, we observe a pronounced nonlinear dependence of the transient reflectivity on fluence that is a marker of the phase-change. To quantify this nonlinearity systematically, we record the ultrafast response at a large range of pump fluences spanning from 5 to 22.5 mJ/cm² (see Supplementary Information S1). The mean transient reflectivity in the asymptotic region was calculated by averaging over the 3 to 4 ps range. The resulting fluence-dependent values are presented in **Fig 2(b)**. A bi-linear behaviour is observed in line with previous studies of $VO_2$[26,29] and other phase change materials[5,41–43] which indicates the crossover from intra-phase to inter-phase changes. For $VO_2$, the bilinear behavior has been used to indicate the phase transition from the monoclinic to the rutile phase. From this data, the fluence threshold is then obtained by fitting a bilinear function and taking the intercept of the steeper linear function (see Methods)[26,29]. At 77 K, we observe a fluence threshold of 17 mJ/cm², in agreement with previous x-ray measurements on similar $VO_2$ samples[28,30].

**Temperature dependence of the fluence threshold**

Next, we evaluate how the fluence threshold depends on the temperature by measuring the fluence-dependent transient reflectivity while scanning the temperature from 25 K to 314 K in steps of 12 K. We probe up to 40K below $T_c$, within the fully insulating monoclinic phase, to avoid any complications from static thermal nucleation of the metallic phase[44] **Fig. 2(c)** shows the critical fluence plotted as a function of temperature. At low temperatures, we observe a nearly constant response. However, as the temperature increases, the fluence threshold decreases monotonically by more than 50%.

We next quantitatively compare the data to the three models discussed previously and illustrated in **Figs. 1(b–d)**. The data are fitted to the generic expression, $F_c = L - U(T)$ where $L$ is treated as a free parameter constant including any relevant latent heat or zero-point energy. We further expand $U(T) = \alpha * P(T)$, where $\alpha$ is a free parameter and $P(T)$ contains the characteristic temperature scaling of each model, i.e. $T^2$ for an electronic transition. The use of a free parameter accounts for uncertainties related to material properties like the absorption length. For the single-mode phonon model, we fix the frequency to the 6 THz mode that dominates the transient response and is often considered to modulate the transition[13,45]. For the thermal model, we use the gapped Debye model and use 250 K and 1200 K as the lower and upper cut offs of the optical phonon spectrum of $VO_2$[17,45].

**Fig 2(c)** illustrates the fits obtained using all three models, along with their corresponding residuals. We find that the electronic heat capacity poorly describes the fluence threshold behavior even neglecting the large quantitative exaggeration. Consequently, we discard this scenario. Similarly, the 6 THz mode scenario markedly disagrees with the observed data, suggesting that this mode is unrelated to the phase transition, in agreement with recent double-pump experiments[29]. Conversely, the gapped Debye model shows good agreement with the experimental data. Our observations suggest that the energy required to trigger the photoinduced structural transition in $VO_2$ cannot be accounted for by the excitation of a single optical phonon mode. Instead, they point to a global lattice response involving a broad range of vibrational modes.

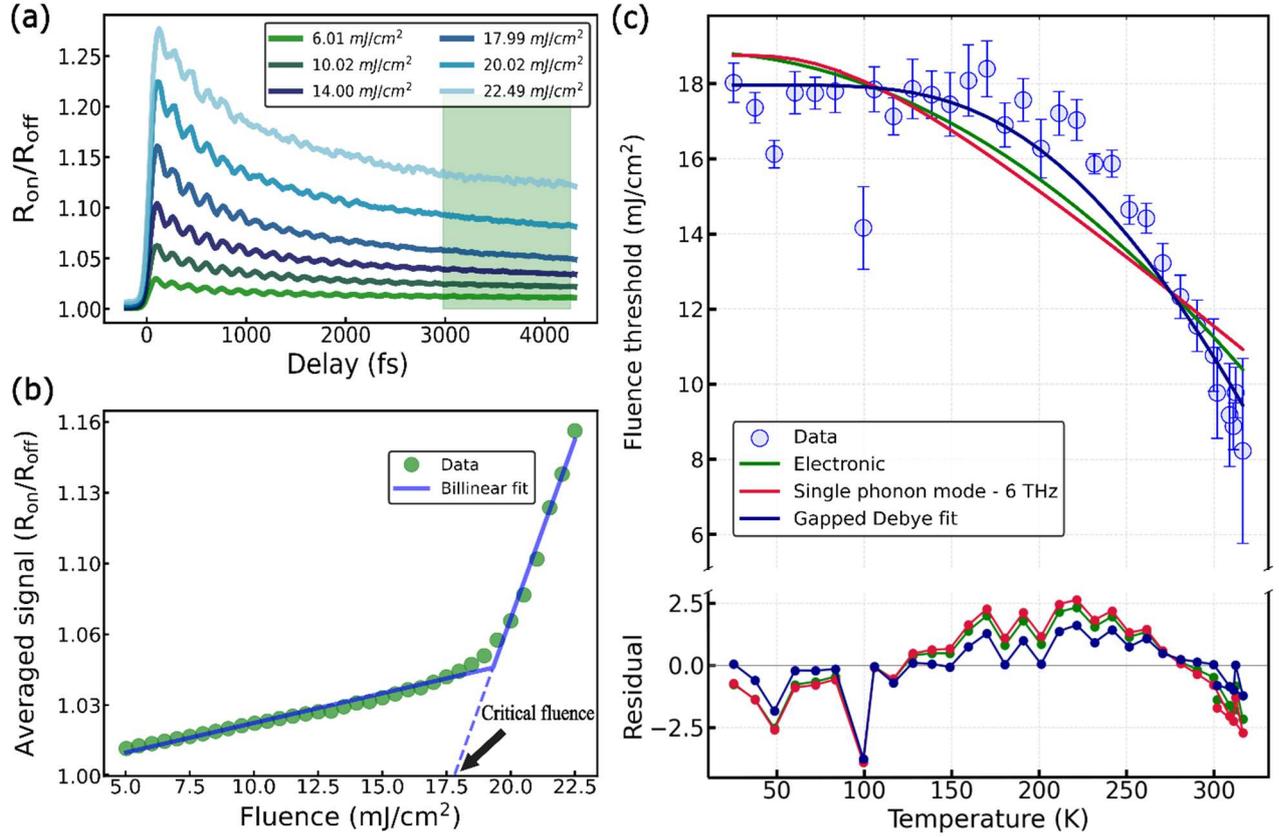

**Fig 2 Temperature Dependent Critical Fluence of VO$_2$:** (a) Pump–probe response of monoclinic VO$_2$ at different fluences using a 700 nm pump and an 1800 nm probe. The data was measured at 77 K. The mean transient reflectivity in the asymptotic region was calculated by averaging over the 3–4 ps interval (green shaded region). (b) Averaged pump–probe signal between 3 and 4 ps at 84 K plotted as a function of fluence (green). The data are fitted with a bilinear function (blue). The critical fluence is defined as the intercept of the second linear component (dashed line: extrapolation of the second linear fit), here around 17.5 mJ/cm². (c) Temperature dependence of the fluence threshold (light blue circles) with the error bars and the fits corresponding to the three studied scenarios: electronic heat capacity (green), single phonon with frequency at 6 THz (red), and gapped Debye fit (dark blue) with the lower cut off of 250 K and an extended upper cutoff temperatures of 1200 K. The residuals corresponding to each fit are shown below in the same colors as their respective fits. Error bars are obtained from the variance and covariances of the bilinear fits (see Methods).

## Temperature dependence of the optical properties and phonon modes

We next consider alternative scenarios that could also explain the observed thermal behavior. Notably, a similar response could conceivably occur if the optical absorption or the coherent phonon cross-section significantly changed with temperature, i.e. the fluence scaling factor $\alpha$ is temperature dependent. To test these scenarios, we examine the temperature dependence of the optical properties and coherent phonon modes and compare these with the results obtained for the critical fluence. High-resolution time-traces were acquired for a range of temperatures from 20 to 330 K at a constant fluence of 6 mJ/cm². An example trace at 100K is shown in **Fig. 3(a),** displaying coherent oscillations resulting from the previously discussed A$_{1g}$ phonon modes. We fit the trace using the following function[29]:

$$\frac{\Delta R}{R} = -0.5\left(1 + erf\left(\frac{t}{t_g}\right)\right)\left(B_e e^{\frac{-t}{t_e}} + A_{ph} e^{\frac{-t}{t_{ph}}} \cos \omega t + C e^{\frac{-t}{t_s}} + E\right) + D, \qquad (1)$$

where $t_g$ is the temporal cross-correlation of the pump and probe pulses, $B_e$ and $C$ are the amplitude of the incoherent contributions, $t_e$ and $t_s$ are the incoherent lifetimes, and $A_{ph}$, $t_{ph}$ are the amplitude of the phonon modes and phonon lifetime respectively, *E* is the long-time asymptotic reflectivity and *D* is a small contribution from static heating. A representative fit along with the corresponding residuals is presented in **Fig. 3(a)**, illustrating that the model accurately captures the observed dynamics; further details are presented in the Methods.

**Fig 3(b)** shows both the amplitude of the coherent phonon modes $A_{ph}$ and the long-time asymptotic reflectivity *E*. Both show a negligible dependence on temperature, and that the optical properties of the VO$_2$ sample remain constant over the full temperature range, and thus the critical fluence dependence cannot be attributed to a change in optical absorption or reflectivity. This is also supported by the negligible temperature-dependence of the negative time-delay reflectivity (Supplementary Information). Furthermore the strength of the coherent lattice oscillations remains essentially unchanged across the studied temperature range, ruling out a variation in the coherent phonon excitation cross section as an origin.

We next examine the possibility of a contribution from mode softening, in analogue to observations in Bi[46] and some charge-density wave systems[47–49]. **Fig. 3(c)** shows the mode frequency $\omega$ as a function of temperature. While the phonon frequency exhibits a clear softening trend, the overall change is less than 5% and seems unlikely to explain the 50% change in critical fluence observed. We further quantify this by fitting the dependence with the same gapped Debye-like function as used to model the critical fluence dependence (purple line in **Fig 3(c)**). The overall agreement is very poor. Instead the data matches well to a quadratic function $A - B*(T)^2$ (blue fit in **Fig. 3(c)**), suggesting that the phonon softening is governed by anharmonic phonon mechanisms distinct from those governing with the phase transition. Finally, we examine the temperature dependence of the decay rate of the A$_{1g}$ phonon mode, as shown in in **Fig. 3(d).** The decay rate exhibits a strong temperature dependence, but one that is uncorrelated with the fluence threshold. However, excellent agreement is found fitting it to an anharmonic decay model (blue curve, details in Methods) in which the optical phonon decays into two acoustic phonons with opposing wave vectors and half the frequency of the optical mode[46,50,51].

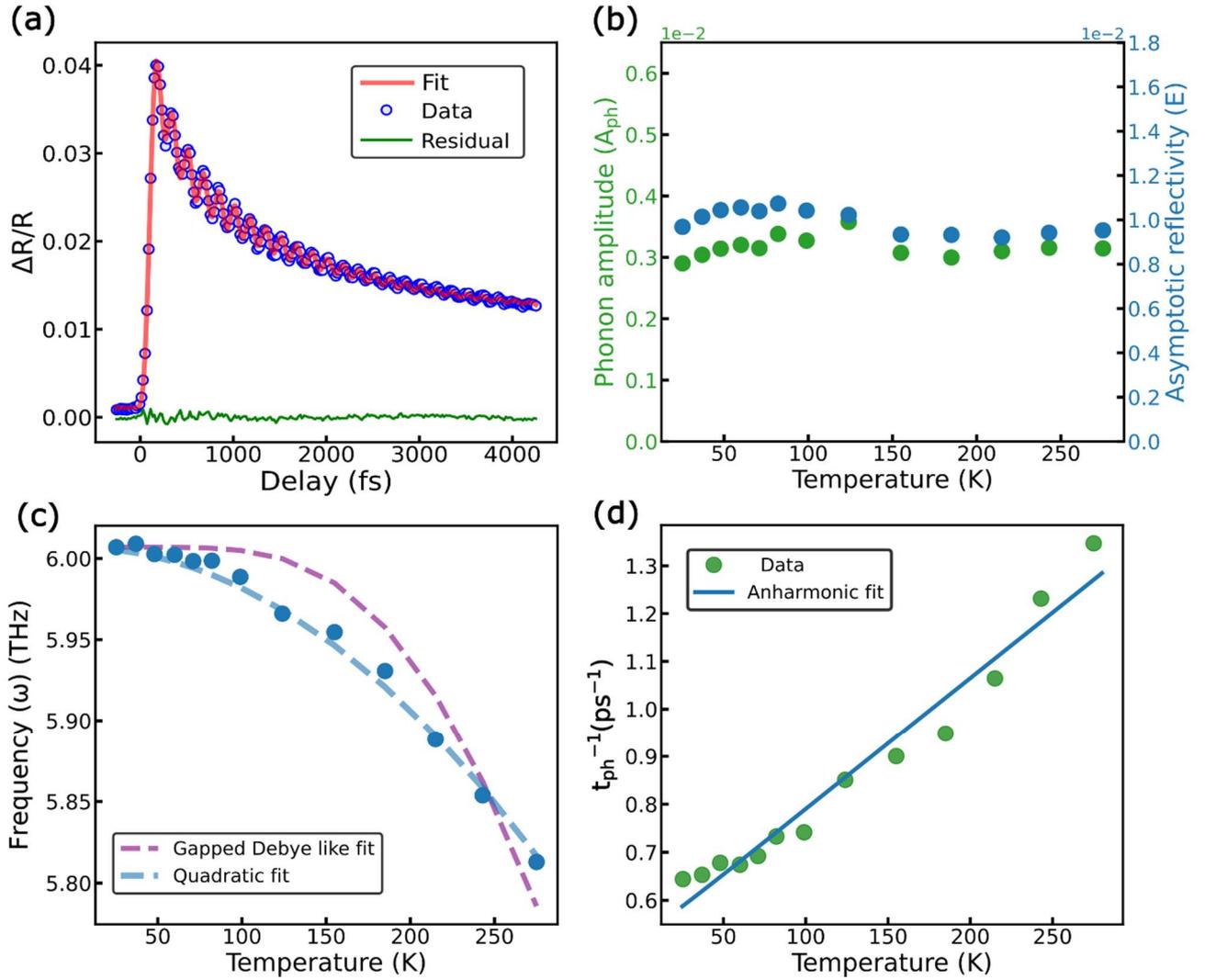

**Fig 3 Coherent Phonon Dynamics**: Pump-probe trace at 100 K with a 700 nm pump and 1800 nm probe at 6 mJ/cm² (blue), showing coherent phonon oscillations. Fit using equation (1) is shown in red and the corresponding residual is plotted in green. (b) Amplitude of the coherent phonon mode ($A_{ph}$ from the fit, green) and the long-time asymptotic reflectivity (parameter $E$ in equation (1), from the fit, blue) as a function of temperature, showing negligible variation across the measured range, indicating largely temperature independent lattice dynamics and optical properties. If the transition were driven by temperature dependent optical changes, E would follow the fluence-threshold trend. (c) Frequency ($\omega$ from the fit) of the coherent phonon mode as a function of temperature. Two models are used to fit the phonon frequency: a gapped Debye-like function (purple), and a quadratic function (blue). The quadratic fit better captures the temperature dependence, suggesting that the phonon behavior is influenced by processes separate from those governing the energy input that induces the phase transition. (d). Decay rate of the phonon mode as a function of the temperature. The solid line (blue) is a fit to the data using an anharmonic decay model (Methods).

## Discussion

Our data shows that the thermal dependence of the photoinduced transition is only well-captured by the gapped-Debye model and that it cannot be attributed to anomalous behaviours of the coherent modes. This shows that the transition proceeds via a disordering or thermally driven mechanism, in which the transition is initiated by a destabilization of the lattice through broadband phonon population. This has several implications.

First, the results indicate that coherence is not a determining factor in driving the structural transition. This is consistent with the total X-ray scattering experiments showing ultrafast order-disorder[8,30] and with the recent double-pump experiments[29] showing coherent lattice mode has no effect on the energy barrier. Our analysis indicates that the small degree of coherence observed across the phase transition at some k-points[8] either has no discernible impact on the mechanics of the transition or may be related to experimental conditions like penetration depth mismatch of the X-rays and optical pump[52].

Second, the results suggest that transient electronic effects influence only the first ~200 fs following photoexcitation, but whether or not the transition is stabilized beyond this time is determined by the energy transfer to the phonon bath. In this framework identifying which phonon bands are most critical to the transition is the central question.

The optical phonons in $VO_2$ range from just below 6 THz up to around 25 THz[17,45]. We find that freely allowing the phonon frequency to vary in the fitting to the single-mode Einstein model yields a best fit at the non-physical frequency of 33.5 THz, though good agreement is also found at 25 THz (Supplementary Information). While neither of these frequencies can be coherently excited by our 12 fs pump pulses, consistent with a disordering pathway, the reasonable fit is indicative that the high-frequency oxygen modes play a key role in the transition.

In agreement with this, we find further variation of the Debye fit parameters shows that excitation of high-energy lattice vibrations is strictly necessary to accurately describe the temperature dependence, with both the non-gapped Debye model and those with a reduced high-frequency cutoff failing to achieve good agreement (see Supplementary Information). Alternative phonon DOS models similarly support that the high-frequency modes are critical; this can be seen directly in the data presented in **Fig. 2c** as the fluence threshold is roughly constant until 200K (17 meV thermal energy), at which point it begins to sharply drop.

The outsized role of the high-frequency phonon bands in the energetics of the ultrafast transition challenges existing models[17,18], which have focused nearly entirely on the low-frequency acoustic bands or the lowest-frequency optical phonons around 6 THz[17,45]. These modes are primarily of vanadium character and begin to be significantly populated well below 200K. We speculate the importance of the high-frequency bands may instead be related to the changes in the phonon density of states through the transition; in the monoclinic phase the phonon cut-off frequency reaches 110 meV, while in the rutile phase this reduces to 80 meV. These changes constitute the largest contribution of phonon entropy to the transition[17], and population of these oxygen character modes may allow the vanadium potential to soften significantly and the dimers to separate in the now flat and highly anharmonic potential[8], even neglecting the corresponding changes in the low-energy acoustic branch. Directly probing the high-energy oxygen phonon modes and their role in renormalizing the lower-energy modes during the ultrafast phase transition could be possible using hard X-ray attosecond pulses now available at X-ray free-electron lasers[53].

We note that our analysis is done on the few-picosecond timescale and therefore our results do not rule out an electronically driven transition and transformations at shorter time delays and lower fluences. Recent few-femtosecond resolution studies have shown that on the sub-100 fs timescale electronic changes can precede structural ones, suggesting the possibility of an electronic-only transition[31]. However, very strong electron-lattice coupling, as observed here and in previous X-ray studies[30], would cause any electronic-only transition to rapidly transfer energy to the lattice and relax to the ground state within the first few-hundred femtoseconds, unless enough energy was provided to also transform the lattice, i.e. heat it above the transition temperature. Any electronic-only transformation should thus be expected to survive only for very short periods of time.

More broadly, our results raise important questions about the limits to creating metastable structural distortions, as disordering may rapidly quench lower-symmetry distortions. However, our results also suggest new routes for controlling ultrafast phase transitions. Specifically, the fact that only a distinct subset of lattice modes are critical for the phase transition, even for a disordering transition like $VO_2$, and that these modes may be distinct from those directly linking the different phases, suggests that photo-engineering of lattice potentials could have broad applicability beyond the coherent-displacive systems considered thusfar[5].

For example, approaches like non-linear phononics[54,55] or cavity confinement[56] could be used not to drive specific modes linked to the desired phase transition, but instead modulate other modes, which in turn control the potential for the order parameter. This strategy is broadly consistent with the observation that photoinduced polaron-like distortions in $VO_2$ can reduce the energy required to initiate the phase transition[29] as these lower frequency distortions also modify the potential landscape and thus likely modulate the renormalization from high-frequency modes observed here. Further experiments combining pre-excitation and thermodynamic approaches could be used to determine exactly how the polarons modulate the potential landscape.

In this context, our thermodynamic approach, which can straightforwardly discern between different types of transitions (i.e. coherent, disordering, lattice, electronic, etc.) without the need for complex multi-messenger experiments and can identify the key modes, is a powerful tool for designing new control strategies for photoinduced phase transitions. Our approach should be broadly applicable for a wide range of systems, especially those in which growing large high-quality crystalline samples required for multi-messenger experiments is challenging. Developing detailed understanding of the ultrafast thermodynamics is central to applying photo-phase-engineering in $VO_2$ and other strongly correlated materials.

## Methods

### Experiments:

A Titanium: Sapphire modelocked laser and amplifier system (Coherent Astrella) produces 7 mJ, sub 100 fs pulses at a central wavelength of 800 nm. Pulses with durations as short as 50 fs can be produced by pumping a non-collinear optical parametric amplifier OPA (TOPAS Prime from Light Conversion), tunable over the near infrared wavelength band (1140 – 3000 nm). These pulses are further compressed via self-phase modulation in a 50 cm hollow core fiber with an inner core diameter of 250 μm, filled with 2.8 bar of statically pumped argon gas. A second non-collinear OPA (TOPAS white, Light Conversion) produces sub-20 fs duration pump pulses tunable across the visible range. Samples are held in a closed-cycle helium cryostat (DE204AF, Advanced Research Systems) that can attain temperatures as low as 15 K and as high as 375 K with sub-10 nm vibrations. Temperature dependent scans were each completed within a few hours, prior to any significant system drift or ice buildup inside the cryostat, which was held below $10^{-6}$ mbar partial pressure.

### Data Fitting:

### Extraction of the Threshold Fluence

The fluence threshold was extracted by fitting a bi-linear function to the change in reflectivity as a function of fluence. The fitting function used is:

$$f(x) = K_1(x - x_0) + y_0 \text{ for } x < x_0 \text{ and } f(x) = K_2(x - x_0) + y_0 \text{ for } x \geq x_0$$

This piecewise linear function has a breakpoint at $x_0$, ensuring continuity at that point $f(x_0) = y_0$. The slopes $K_1$ and $K_2$ describe the linear behavior of the reflectivity response below and above the critical fluence, respectively. The fluence threshold is defined as the intercept of the second linear function with the baseline reflectivity i.e. $(1 - y_0 + K_2 x_0)/K_2$. The uncertainty in the extracted fluence threshold ($\sigma_{FT}$) was evaluated using standard error propagation, taking into account the variances and covariances of the fit parameters from the covariance matrix of the bilinear fit, as expressed by:

$$\sigma_{FT} = FT \left[ \frac{Var(x_0)}{x_0^2} + \frac{Var(y_0)}{y_0^2} + \frac{Var(K_2)}{K_2^2} - 2 \frac{Cov(x_0, K_2) + Cov(y_0, K_2) - Cov(x_0, y_0)}{x_0, y_0, K_2} \right]^{0.5}$$

where $FT$ denotes the extracted fluence threshold, and $Var$ and $Cov$ represent the variances and covariances of the corresponding fit parameters, respectively. The extracted thresholds and their corresponding errors as a function of temperature are shown in **Fig. 2(c)** of the main text.

### Time-Resolved Fitting of Coherent Phonon Dynamics

High-resolution transient reflectivity time traces, exhibiting coherent phonon oscillations as shown in **Fig. 2(a)** of the main text, were analyzed using a multi-component fitting model (equation (1) in the main text) that accounts for both coherent and incoherent contributions. The fitting revealed two distinct incoherent exponential decay components: a fast decay with a lifetime of approximately 230 fs and a slower decay of about 1600 fs. The previously discussed 10 THz mode was found to be too weak to be fit reliably across all datasets; thus, only one mode is considered in the fitting of the phonon dynamics. This restriction was found not to affect the quality of the overall fit or the obtained trends for the 6 THz mode.

**Anharmonic Decay Model**

The decay rate for the anharmonic phonon decay mechanism is given as[50]

$$t_{ph}^{-1} = t_{p0}^{-1}\left[1 + \frac{2}{\exp\left[(\hbar\Omega/2)/k_B T\right] - 1}\right] + C, \qquad (2)$$

where $t_{p0}^{-1}$ is the fitting parameter which describes the effective anharmonicity, $C$ describes the phonon-electron or phonon-defect collision rate, and $\Omega$=6 THz is the phonon frequency. This yields a phonon-phonon scattering of $0.03\ ps^{-1}$ and an electron-phonon/electron-defect scattering rate of $0.5\ ps^{-1}$, suggesting the phonon decay is dominated by electron-phonon coupling, in agreement with other recent works on $VO_2$[29,30].

**Data availability**

All data shown in this study are included in this published article (and its Supplementary Information files). Source data are provided with this paper.

**Competing Interests**

The authors declare no competing interests.

**Acknowledgements**

This work was funded by the Spanish AIE (projects PID2022-137817NA-I00 and EUR2022-134052) and the Comunidad de Madrid (project TEC-2024/TEC-380 "Mag4TIC"). Funded by the European Union (ERC, KnotSeen, 101163311). Views and opinions expressed are however those of the author(s) only and do not necessarily reflect those of the European Union or the European Research Council. Neither the European Union nor the granting authority can be held responsible for them. ASJ acknowledges the support of the Ramón y Cajal Program (Grant RYC2021-032392-I), SB acknowledges FPI fellowship PRE2022-103187. IMDEA Nanociencia acknowledges support from the "Severo Ochoa" Programme for Centers of Excellence in R&D (MICIN, CEX2020-001039-S). SEW acknowledges support from Novo Nordisk Foundation (NNF23OC0084990—Harnessing Dynamic Disorder for Efficient Material Control (HD-control)). EP thanks the support of CNRS and the French Agence Nationale de la Recherche (ANR), under grant ANR-22-CPJ2-0053-01

**Author Contributions**

SEW conceived the study, JS grew the VO$_2$ sample, ASJ and EP performed the experiments, ASJ and SB analyzed the data, all authors contributed to the interpretation of the results and writing of the article.

# Supplementary Information

## Temperature Dependent Pump-Probe Measurements

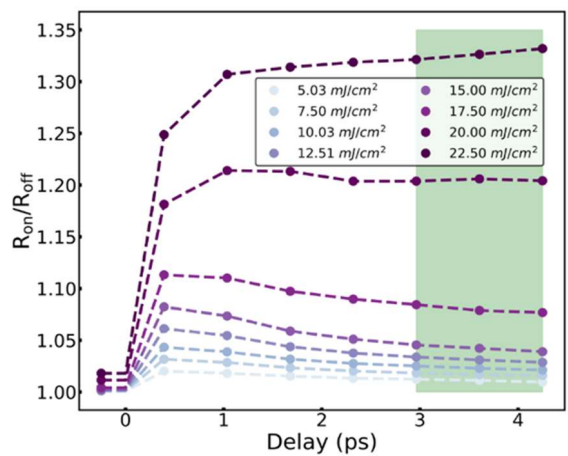

**Figure S1:** Transient reflectivity measurements for different fluences with a 700 nm pump and an 1800 nm probe were carried at a range of temperatures. The result shown is for the measurement carried out at 235 K. Rapid pump-probe measurements with only 8 time points were recorded in order to acquire the full range of data within a few hours prior to any significant system drift or ice buildup inside the cryostat. Only few fluence values are shown in the plot for clarity. The dashed lines are guides for the eye. The mean transient reflectivity in the asymptotic region was calculated by averaging over the 3–4 ps interval (green shaded region). This average value is plotted as a function of fluence in **Fig 2(b)** of the main text.

## Thermal transition modelling of the fluence threshold

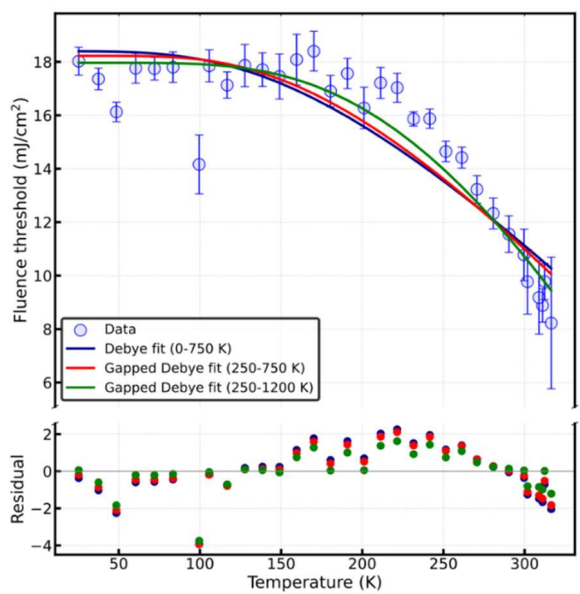

**Figure S2**: Temperature dependence of the fluence threshold (light blue circles) with the error bars and the model fits: Debye fit (dark blue), gapped Debye fit (red) with the lower and upper cutoff temperatures of 250 K and 750 K, respectively and gapped Debye fit (green) with the lower cut off of 250 K and an extended upper cutoff temperatures of 1200 K. The residuals corresponding to each fit are shown below in the same colors as their respective fits.

The temperature dependent fluence threshold are fitted with the same expression $F_c = L - U(T)$ as explained in the main text. Three cases are considered here: (1) a full Debye model with a Debye temperature $T_D = 750$ K, (2) a gapped Debye model with a lower cutoff of 250 K and an upper cutoff of 750 K, and (3) the gapped Debye model with a lower cutoff of 250 K and an upper cutoff of 1200K discussed in the main text. The value of 750K is selected to be in agreement with previous thermal measurements of the heat capacity in $VO_2$[36,37], though conventional thermal measurements primarily probe the acoustic phonon branch which is not relevant for the ultrafast transition. Only the third model allows higher-energy optical phonons to contribute.

Figure S2 shows the three models compared to the temperature dependent fluence threshold data from the main text. From the results, it is evident that both the full Debye fit, which accounts for the occupation of all phonon states up to the Debye temperature (750 K), and the gapped Debye fit with an upper cutoff fixed at 750 K, fail to accurately reproduce the experimental fluence-threshold behavior. This discrepancy indicates that simply considering energy distributed among the low- and mid-frequency lattice modes is insufficient to capture the observed dynamics. In particular, the inability of these models to describe the data suggests that excitation of the high-frequency optical phonon modes is essential for effectively driving the structural phase transition in $VO_2$. The gapped Debye fit with the same lower cutoff of 250 K but an extended upper cutoff of 1200 K captures the experimental behavior more accurately. This finding indicates that the photoinduced transition depends especially on excitation of high-energy lattice vibrations.

**Additional single-mode fits**

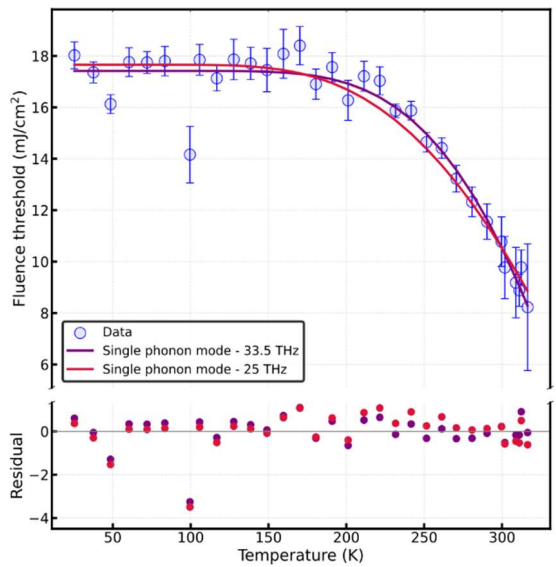

**Figure S3**: Temperature dependence of the fluence threshold (light blue circles) with the error bars and the model fits: Einstein model with 33.5 THz mode (purple) and 25 THz (red). The residuals corresponding to each fit are shown below in the same colors as their respective fits.

Allowing the mode frequency in the Einstein model to vary freely yields a best fit at 33.5 THz, as shown in Figure S3. This frequency is unphysical, as the cutoff frequency in VO2 is known to be around 25 THz. Restricting to this value again yields good agreement with the experimental data, also shown in Figure S3.

# Temperature-dependence of the negative time-delay reflectivity and static heating

To ensure that the optical response of the VO$_2$ sample remains constant across the entire temperature range, and to confirm that the observed variation in the critical fluence is not influenced by temperature-dependent changes in optical absorption or reflectivity, we analyzed the temperature dependence of the negative time-delay reflectivity. This region, serves as a sensitive probe of any static or cumulative heating effects that could modify the baseline reflectivity. The resulting map, shown Figure S4a, presents the ratio $\frac{R(pump\ off)}{R(pump\ off\ at\ 25\ K)}$ at negative time delay for different temperatures and fluences. Across the entire range investigated, no systematic trends or anomalies were observed, indicating that the optical constants of VO$_2$ remain effectively unchanged below the transition temperature. In addition, the static heating contribution, represented by the parameter D in the fitting equation (equation (1) in the main text) used to model the temporal coherent dynamics, is also plotted in Figure S4b. The negligible variation of D with temperature corroborates the conclusion drawn from the negative time-delay analysis.

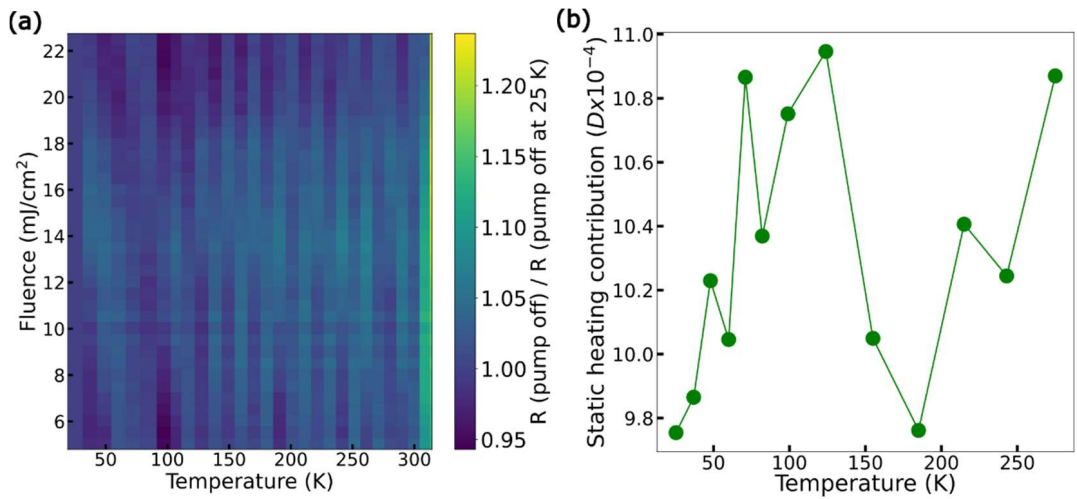

**Figure S4**: (a) Two-dimensional map showing the ratio $R(pump\ off)/R(pump\ off\ at\ 25\ K)$ at negative time delay, plotted as a function of temperature and fluence. The absence of systematic variations across the measured range indicates that the static reflectivity of VO$_2$ remains unchanged below the phase transition. (b) Temperature dependence of the static heating contribution D, obtained from fitting the transient reflectivity dynamics using the model described in the main text.

Page **18** of 18